\documentclass[12pt]{article}
\usepackage[left=2.50cm, right=2.50cm, top=2.50cm, bottom=2.50cm]{geometry}
\usepackage[utf8]{inputenc}
\usepackage[english]{babel}
\usepackage{physics}
\usepackage{hyperref}
\usepackage{amsthm}
\usepackage{mathtools}
\usepackage{graphicx}
\usepackage{changepage}
\usepackage{verbatim}
\usepackage{empheq}
\usepackage{xcolor}
\usepackage{tikz}
\usepackage{multirow}
\usetikzlibrary{tikzmark}
\numberwithin{equation}{section}
\hypersetup{hidelinks}

\newlength{\bibitemsep}
\newlength{\bibparskip}
\let\oldthebibliography\thebibliography
\renewcommand\thebibliography[1]{%
  \oldthebibliography{#1}%
  \setlength{\parskip}{\bibitemsep}%
  \setlength{\itemsep}{\bibparskip}%
}
\begin{document}
\noindent

{\bf
{\Large 
Fierz-Pauli theory reloaded: from a theory of a symmetric tensor field  to linearized massive gravity}
}

\vspace{.5cm}
\hrule

\vspace{1cm}

\large
\noindent
{\bf Giulio Gambuti$^{1,a}$}
and
{\bf Nicola Maggiore$^{2,3,b}$}\\

\par
\small
\noindent$^1$ Rudolf Peierls Centre for Theoretical Physics, Oxford University, U.K.
\smallskip

\noindent$^2$ Dipartimento di Fisica, Universit\`a di Genova, Italy.
\smallskip

\noindent$^3$ Istituto Nazionale di Fisica Nucleare - Sezione di Genova, Italy.

\setcounter{footnote}{0}

\noindent

{

\vspace{1cm}

\noindent
{\tt Abstract~:}
Modifying gravity at large distances by means of a massive graviton may explain the observed acceleration of the Universe without Dark Energy. The standard paradigm for Massive Gravity is the Fierz-Pauli theory, which, nonetheless, displays well known flaws in its massless limit. The most serious one is represented by the vDVZ discontinuity, which consists in a disagreement between the massless limit of the Fierz-Pauli theory and General Relativity. Our approach is based on a field-theoretical treatment of Massive Gravity: General Relativity, in the weak field approximation, is treated as a gauge theory of a symmetric rank-2 tensor field. This leads us to propose an alternative theory of linearized Massive Gravity, describing five degrees of freedom of the graviton, with a good massless limit, without vDVZ discontinuity, and depending on one mass parameter only, in agreement with the Fierz-Pauli theory.
\vspace{.5cm}

\vspace{\fill}
{\tt Keywords:} Linearized General Relativity; Massive Gravity; Fierz-Pauli Theory; Large Distances Modified Gravity.
\vspace{1cm}

\hrule
\noindent{\tt E-mail:
$^a$giulio.gambuti@new.ox.ac.uk,
$^b$nicola.maggiore@ge.infn.it.}
\newpage

\section{Introduction}

\subsubsection*{Motivations}

It is an observational fact that the Universe is expanding at an accelerated rate \cite{Riess:1998cb,Perlmutter:1998np}. Within the theory of General Relativity (GR), the cosmological constant  $\Lambda$ explains this phenomenon. In a picture of the Universe seen as a perfect fluid, the cosmological constant gives rise to an accelerated expansion by acting as a constant energy density (called Dark Energy) $\rho \sim \Lambda M_P^2$ providing a negative pressure. A fine tuning of the constant $\Lambda$ enables us to match the expansion predicted by GR to the one which is observed. The estimate from observations is $\frac{\Lambda}{M_P^2} \sim 10^{-65}$ \cite{Ade:2015xua}, where $M_P$ is the Planck mass, while the quantum field theoretical prediction on the cosmological constant seen as the vacuum energy density gives $\frac{\Lambda}{M_P^2} \sim 6 \times 10^{54}$. Unfortunately, these two estimates disagree by about 120 orders of magnitude. This huge tension is known as the cosmological constant problem \cite{Weinberg:1988cp}, which motivates alternative descriptions for Dark Energy, with the request that any cosmological model should reproduce an Universe which, at our epoch, is almost perfectly flat and filled by matter and DE in the ratio of about 3/7, where the DE is effectively approximated by a constant \cite{Wetterich:1987fm,Caldwell:1997ii,ArmendarizPicon:2000dh,Damour:1994zq,Amendola:1999er,Bonici:2018qli}.
Modifying gravity at large distances by means of a massive graviton is a possible solution to the cosmological constant problem \cite{Hinterbichler:2011tt}. In fact, the Yukawa potential for a massive field at large distances goes like $\propto \frac{1}{r} e^{- \alpha m r}$, where $m$ is the mass of the field and $\alpha$ a dimensional constant. At scales comparable to $\frac{1}{\alpha m}$, the exponential factor suppresses the potential and the strength of interactions as well. This is the reason why long-range forces are associated with massless bosons and short-range forces with massive bosons. 
By means of a Yukawa-like potential, the gravitational effect of the vacuum energy density is exponentially suppressed at large scales, thus explaining the disagreement between the quantum field theory calculation and the observed cosmological value. 
Clearly, this exponential suppression of long-range gravitational interactions is constrained by experimental evidence and, consequently,  the mass of the graviton is subject to restrictive upper limits.  
Another way of understanding this point is to remember that a mass acts as a momentum space cutoff of interactions, effectively damping the effect of low frequency sources. The vacuum energy density can be considered constant, therefore its contribution to the expansion of the Universe is greatly reduced by the introduction of a mass for the graviton, again confirming that in a picture where the graviton is massive the cosmological constant problem might be fixed by a suitable graviton mass.
Theories in which the graviton is massive are referred to as Massive Gravity (MG) theories.
We nonetheless remark that the introduction of a mass for the graviton as a 
potential solution of the cosmological constant problem
was a hope more than a
decade ago, but early
experience with nonlinear MG  (see, for example \cite{Hassan:2011vm}) showed that this expectation is not realized in the
nonlinear theory where a cosmological constant bends spacetime in the
same way as in GR. While a large vacuum energy can be canceled by a
MG contribution, this still requires fine tuning and does
not amount to a degravitation mechanism.\footnote{ We thank the referee for this remark.}
In some sense MG is a substantial modification of GR, even if the mass of the graviton is extremely close to zero. In fact, when adding a mass, the two degrees of freedom (DOF) carried by the massless boson become $2S+1$ massive DOF. For the graviton, which is the spin $S=2$ representation of the Poincar\'e group, the two massless DOF become five massive ones. 

\subsubsection*{Phenomenological Limits}

As previously said, long range forces are usually carried by massless bosons because the Yukawa exponential factor limits the range of interactions mediated by massive particles. Nevertheless, choosing a sufficiently small value of the graviton mass, the interactions at scales of the observable Universe remain essentially identical to those of ordinary gravity, while only interactions at greater scales are affected. Indeed, the observed accelerated rate of the Universe gives an upper limit to the mass of the graviton, which in \cite{DAmico:2011eto} is shown to be of order $10^{-34} \; \frac{eV}{c^2}$, roughly an order of magnitude smaller than the Hubble parameter at our epoch.  With such a limit, for interactions at distances much smaller than the radius of the observable Universe (which is estimated to be about $46.6 \times 10^9 \: \mathrm{ly}$ \cite{Gott:2003pf}), the Yukawa exponential drops to one, and the classical Newton potential $\propto \frac{1}{r}$ is recovered, as desired. Still, this cosmological upper limit becomes irrelevant when the mass is acquired through the condensation of some additional scalar field (see \cite{DAmico:2011eto}).
Another constraint on the mass of the graviton was recently obtained by the LIGO-Virgo collaboration through the analysis of binary black holes merger signals, measuring the phase shift between components of different frequencies of the gravitational waves detected by the interferometers \cite{LIGOScientific:2019fpa}. In this way, the upper limit for the mass of the graviton is estimated to be $4.7 \times 10^{-23} \; \frac{eV}{c^2}$, which is several orders of magnitude higher than the limit derived in \cite{DAmico:2011eto} by cosmological considerations.

\subsubsection*{State of the Art}

In 1939 Fierz and Pauli (FP)  proposed a relativistic theory for a massive particle with arbitrary spin $f$, described by a symmetric rank-$f$ tensor field \cite{Fierz:1939ix}. They showed that ``in the  particular case of spin two, rest-mass zero, the equations agree in the force-free case with Einstein's equations for gravitational waves in GR in first approximation; the corresponding group of transformations arises from the infinitesimal coordinate transformations''. This is what is known as the FP theory, which is considered as the standard paradigm for Linearized MG (LMG).
Much later, it has been realized by van Dam, Veltman and Zakharov \cite{vanDam:1970vg,Zakharov:1970cc} that the massless limit of the FP theory exhibits a discontinuity with GR, known as vDVZ discontinuity.
In particular, in \cite{Rubakov:2008nh} it is shown that the interaction of light with a massive body, like a star for instance, is 25\% smaller in the massless limit of the FP theory than in GR. Consequently, the angle of deviation of light is also different by 25\% in the two theories, a difference which can be observed by means of the time delay in the gravitational lensing phenomenon.  The problems of the FP theory at small scales can be solved by the  Vainshtein mechanism \cite{Vainshtein:1972sx}, which nonetheless requires a nonlinear modification of the FP theory.
In principle, this measurable disagreement between GR and the FP theory of LMG would allow us to distinguish between a strictly vanishing mass and a very small one. Such a tension is crucial, since GR has been extensively tested and as a consequence we might rule out the FP theory as a theory of LMG.
It was also pointed out in \cite{Boulware:1973my} that in order for Linearized Gravity (LG) to describe a pure spin two system, it is necessary to use a particular FP mass term, referred to as the FP tuning. Otherwise, there will be admixtures of lower spin, in general with negative-energy, called Boulware-Deser ghost.  Moreover, Boulware and Deser argued that the ghost will reappear if nonlinear extension of LG are considered. Some possibilities to circumvent these problems have been proposed in \cite{deRham:2010ik,deRham:2010kj,Hassan:2011hr}. 
A detailed review on MG can be found in \cite{deRham:2014zqa}.

\subsubsection*{Strategy}

The approach adopted in this paper is based on a field-theoretical treatment of LG. We consider LG as a gauge theory of a symmetric rank-2 tensor field, the gauge symmetry being the infinitesimal diffeomorphism invariance \cite{Blasi:2015lrg}. Before adding a mass term to the action, which breaks the gauge symmetry, the massless theory should be well defined, in the sense that a well defined partition function must exist. In gauge field theory this is achieved by gauge fixing the action.  This procedure (gauge fixing the massless action first, adding a breaking mass term after) leads to a gauge theory of LMG
describing five DOF for the massive graviton, with a good massless limit and without the vDVZ discontinuity. In the FP theory, instead, the mass term is added directly to the invariant action and effectively plays the double and unnatural role of both mass term {\it and} gauge fixing. Basically, this is the reason why the FP theory displays an ill-defined massless limit, which shows itself in a divergent propagator and, physically, in the vDVZ discontinuity with GR. We stress that any theory of LMG should display a good massless limit since the phenomenological limits on the mass of the graviton are such that the mass term must be seen as a small perturbation of massless LG and, equivalently, as a small breaking of diffeomorphism invariance.

\subsubsection*{Summary}

This paper is organized as follows: in Section 2 the gauge fixed massive theory is presented. We also review the FP theory and the problems related to its massless limit which motivate this work, and we remark that, curiously, the original 1939 paper by Fierz and Pauli does not exactly describe what is generally known as the FP theory. In Section 3 we show, in a gauge independent way, that the propagating DOF of the ten components of the rank-2 symmetric tensor describing the massive graviton are indeed five, and depend on one mass parameter only, in agreement with the FP theory. In Section 4 we compute the propagator of the theory, we show that its massless limit is regular and that the same observable which is used to unveil the vDVZ discontinuity presents an unique pole in the same mass parameter which characterizes the five DOF of the theory. In Section 5 we show the absence of the vDVZ discontinuity, for any gauge choice. Our results are summarized and discussed in the concluding Section 6.\\

List of acronyms: 
DOF: Degree(s) of Freedom;
EOM: Equation(s) of Motion;
FP: Fierz-Pauli;
GR: General Relativity;
LG: Linearized Gravity;
LMG: Linearized Massive Gravity;
MG: Massive Gravity;
vDVZ: van Dam, Veltman and Zacharov.

\section{The massive action}

The weak field expansion of GR around the flat Minkowskian background $\eta_{\mu \nu} = diag(-1,1,1,1)$ is given by the LG action

\begin{equation}\label{SLG}
        S_{LG}[h]= \int \mathrm{d^4 x} \; \;  \Big[ \; \frac{1}{2} h \partial^2 h \;  - \; h_{\mu \nu } \partial^\mu \partial^\nu h \; - \;\frac{1}{2} h^{\mu \nu} \partial^2 h_{\mu \nu}\;+ \;h^{\mu \nu} \partial_\nu \partial^\rho h_{\mu \rho} \; \Big]\; ,
\end{equation}
where $h_{\mu \nu}(x)$ is a symmetric rank-2 tensor field representing the graviton, and \mbox{$h(x)\equiv\eta_{\mu\nu}h^{\mu\nu}(x)$} is its trace.
The action $S_{LG}[h]$ \eqref{SLG} is the most general functional invariant under the infinitesimal diffeomorphism transformation
\begin{equation}\label{gaugesym}
\delta h_{\mu \nu}(x) = \partial_\mu \xi_\nu(x) + \partial_\nu \xi_\mu(x)\; ,
\end{equation}
where $\xi_\mu(x)$ is a local vector parameter. The transformation \eqref{gaugesym} represents the gauge symmetry of the action $S_{LG}$ \eqref{SLG}.

\subsection{The Fierz-Pauli theory}

The most general mass term which can be added to the invariant action $S_{LG}$ \eqref{SLG}, respecting Lorentz invariance and power counting,  is  
\begin{equation}\label{Sm}
S_m[h;m_1^2,m_2^2] =\frac{1}{2}\int \mathrm{d}^4x \ (m_1^2 h_{\mu \nu} h^{\mu \nu} +m_2^2h^2 )\; ,
\end{equation}
where $m_1^2$ and $m_2^2$ are massive parameters. The presence of a mass term breaks the diffeomorphism invariance \eqref{gaugesym}, as usual in any gauge field theory. It can be shown (see for instance \cite{deRham:2014zqa}) that the action 
\begin{equation}\label{S}
S =S_{LG}+S_m\; .
\end{equation}
describes the propagation of five DOF only if 
\begin{equation} 
m_1^2+m_2^2=0\ ,
\label{FPT}\end{equation}
otherwise a sixth ghost mode with negative energy appears \cite{Boulware:1973my}, and the theory does not describe a massive graviton.
The choice \eqref{FPT} is generally referred to as FP tuning, and the FP theory is defined by the action
\begin{equation}\label{SFP}
S_{FP}[h;m_1^2] \equiv S_{LG}[h] + S_m[h;m_1^2,-m_1^2]\; .
\end{equation}
Following \cite{Hinterbichler:2011tt}, we now show that the theory described by the action $S_{FP}$ \eqref{SFP} does indeed display five DOF.
The Equations of Motion (EOM) obtained from \eqref{SFP} read
\begin{equation}\label{EOMs}
\frac{\delta S_{FP}}{\delta h^{\mu \nu}} = \partial^2 h_{\mu \nu} - \partial_{\alpha} \partial_\mu {h^\alpha}_\nu - \partial_{\alpha} \partial_\nu {h^\alpha}_\mu + \eta_{\mu \nu} \partial_\alpha \partial_\beta h^{\alpha \beta} +  \partial_\mu \partial_\nu h - \eta_{\mu \nu} \partial^2 h
 - m_1^2 (h_{\mu \nu} - \eta_{\mu \nu}h) = 0\; ,
\end{equation}
which, saturated with $\partial^\mu$, yield the constraint
\begin{equation}\label{constr}
\partial^\mu h_{\mu \nu} - \partial_\nu h = 0\; .
\end{equation}
Plugging \eqref{constr} into \eqref{EOMs} we get 
\begin{equation}\label{EOMs2}
\partial^2 h_{\mu \nu} - \partial_\mu \partial_\nu h - m_1^2 (h_{\mu \nu} - \eta_{\mu \nu}h)= 0\; .
\end{equation}

Saturating \eqref{EOMs2} with $\eta^{\mu \nu}$ we find 
\begin{equation}
h=0\; ,
\end{equation}
which, together with \eqref{constr}, implies 
\begin{equation}\label{transversality}
\partial^\mu h_{\mu \nu}= 0\; .
\end{equation}
Therefore, the EOM \eqref{EOMs} imply the following set of equations
\begin{align}
&(\partial^2 - m_1^2)h_{\mu \nu}(x)=0 \;   \label{K-G}\\[8pt]
&\partial^\mu h_{\mu \nu}(x)=0 \;  \label{C1}\\[8pt]
&h(x)=0\;. \label{C2}
\end{align}

Eq. \eqref{K-G} is the Klein-Gordon equation for the field $h_{\mu \nu}(x)$, while \eqref{C1} and \eqref{C2} represent five constraints (transversality and tracelessness) which reduce the ten independent components of $h_{\mu \nu}$ to five. These five components carry the five massive DOF of the graviton.

\subsection{Problems with the Fierz-Pauli theory}

The propagator of the FP theory \eqref{SFP} is 
\begin{equation} \label{FPprop}
 G^{FP}_{\mu \nu, \alpha \beta}(p) = \frac{2}{p^2 + m_1^2} \left[ \frac{1}{2}(P_{\mu \alpha} P_{\nu \beta} + P_{\nu \alpha} P_{\mu \beta}) - \frac{1}{3} P_{\mu \nu} P_{\alpha \beta} \right]\; ,
\end{equation}
where $P_{\mu \nu}$ is the transverse massive projector defined as
\begin{equation}
P_{\mu \nu}=  \eta_{\mu \nu} + \frac{p_\mu p_\nu}{m_1^2}\; .
\label{FPtransv}\end{equation}
A crucial remark is that the propagator \eqref{FPprop} exists only thanks to the presence of the mass term \eqref{Sm} on the FP point \eqref{FPT}. As a consequence of this fact, it is apparent from  \eqref{FPtransv} that the FP theory has a divergent massless limit. 
Moreover, the massless limit of the FP theory is flawed by the vDVZ discontinuity \cite{vanDam:1970vg,Zakharov:1970cc}, which basically consists in the fact that the correlator involving two energy-momentum tensors, computed in the FP theory in the limit $m_1^2 \rightarrow 0$, does not match the GR prediction. We sketch here the proof (for details see \cite{Hinterbichler:2011tt,Rubakov:2008nh}). 
The gravitational interaction between two non relativistic energy-momentum tensors ${T_{\mu \nu}^{(1)}}$ and ${T^{(2)}}_{\mu\nu}$ (which are conserved, $i.e.$ $p^\nu \tilde{T}_{\mu \nu}^{(1)} =p^\nu \tilde{T}_{\mu \nu}^{(2)}= 0$) is described by the introduction in the action of an interaction term of the type
\begin{equation}\label{int term} 
S_{int} = \lambda \int \mathrm{d}^4 x \;  h_{\mu \nu} T^{\mu \nu}\; ,
\end{equation}
where $T_{\mu \nu}$ denotes a generic energy-momentum tensor, coupled to $h_{\mu\nu}(x)$ through a constant $\lambda$, which we call $\lambda_{LG}$ and $\lambda_{FP}$
for LG and FP theory, respectively.
Therefore, the interaction strength between ${T_{\mu \nu}^{(1)}}$ and ${T_{\mu \nu}^{(2)}}$ can be computed by contraction with the propagator of the graviton. In LG, the propagator $G_{LG}^{\mu \nu, \alpha\beta}$ is obtained from the LG action \eqref{SLG} after a gauge fixing, as done in \cite{Hinterbichler:2011tt}. The FP propagator $G_{FP}^{\mu \nu, \alpha \beta}$, on the other hand, is given by \eqref{FPprop}.
In the non relativistic limit, only the 00-components of the energy-momentum tensors are non negligible. Two cases are considered: in the first, ${T_{\mu \nu}^{(1)}}$ and ${T_{\mu \nu}^{(2)}}$ are associated with massive objects and therefore have non vanishing trace. In the second case, ${T_{\mu \nu}^{(1)}}$ still has a non vanishing trace whereas ${T_{\mu \nu}^{(2)}}$ is traceless, representing, for instance, electromagnetic radiation ($e.g.$ light). 
Concerning the first case, the interaction strength in LG is
\begin{equation}\label{INT1}
\lambda^2_{LG}\: \tilde{T}^{(1)}_{\mu \nu} G_{LG}^{\mu \nu, \alpha \beta} \tilde{T}^{(2)}_{\alpha \beta} = \lambda^2_{LG}\: \tilde{T}^{(1)}_{00} \tilde{T}^{(2)}_{00} \frac{1}{p^2}\; ,
\end{equation}
while, in the second, $i.e.$ when ${T_{\mu \nu}^{(2)}}$ is traceless, we have
\begin{equation}\label{INT2}
\lambda^2_{LG}\: \tilde{T}_{\mu \nu}^{(1)} G_{LG}^{\mu \nu, \alpha \beta} \tilde{T}^{(2)}_{\alpha \beta} = \lambda^2_{LG}\: \tilde{T}^{(1)}_{00} \tilde{T}^{(2)}_{00} \frac{2}{p^2}\; .
\end{equation}
On the other hand, the interaction strengths corresponding to \eqref{INT1} and \eqref{INT2} obtained using the FP propagator \eqref{FPprop}, in the massless limit $m_1 \rightarrow 0$ are, respectively
\begin{equation}\label{INT1FP}
\lambda^2_{FP}\: \tilde{T}_{\mu \nu}^{(1)} G_{FP}^{\mu \nu, \alpha \beta} \tilde{T}^{(2)}_{\alpha \beta} = \frac{4}{3} \lambda^2_{FP}\: \tilde{T}^{(1)}_{00} \tilde{T}^{(2)}_{00} \frac{1}{p^2}
\end{equation}
and
\begin{equation}\label{INT2FP}
\lambda^2_{FP} \:\tilde{T}_{\mu \nu}^{(1)} G_{FP}^{\mu \nu, \alpha \beta} \tilde{T}^{(2)}_{\alpha \beta} = \lambda^2_{FP} \:\tilde{T}_{00}^{(1)} \tilde{T}_{00}^{(2)} \frac{2}{p^2}\; .
\end{equation}
The vDVZ discontinuity arises from the observation that it is not possible to choose an unique FP coupling $\lambda_{FP}$ which matches both \eqref{INT1} with \eqref{INT1FP} and \eqref{INT2} with \eqref{INT2FP}: the value $\lambda^2_{FP} = \frac{3}{4}\lambda^2_{LG}$ matches the first pair but not the second.
The interaction strength between a massive body and an electromagnetic wave in the massless limit of the FP theory turns out to be $\frac{3}{4}$ of the LG prediction, hence the discontinuity.

\subsection{Back to the origins}

In \cite{Fierz:1939zz} Fierz studied the general relativistic theory of force-free particles with any (integer or half-integer) spin $f$, described by symmetric ``world'' tensors 
$A_{ik...l}(x)$\footnote{ In this subsection we respectfully maintain  the notation used in the original paper by Fierz and Pauli in 1939. In particular, the indices $(i, j)$ refer to spacetime, contrarily to the rest of the paper, where greek letters are used.}. The starting point in \cite{Fierz:1939zz}. The starting point in \cite{Fierz:1939zz} is the request that the fields $A_{ik...l}(x)$ satisfy the massive wave equation (which Fierz called of the  ``Schr\"{o}dinger-Gordon type'')
\begin{equation}
\partial^2 A_{ik...l} = \kappa^2 A_{ik...l}\ ,
\label{SGeq}\end{equation}
where $\kappa$ is a constant with the dimension of the inverse of a length (to which corresponds the mass $m=\frac{\hbar\kappa}{c}$). In addition, the fields $A_{ik...l}(x)$ are asked to satisfy the additional constraints (called ``secondary conditions'')
\begin{eqnarray}
A_{ii...l} &=& 0 \label{trace}\\
\partial_i A_{ik...l} &=& 0\ .\label{transv}
\end{eqnarray}
As explained in \cite{Fierz:1939zz}, the secondary conditions \eqref{trace} and \eqref{transv} are introduced in order to ensure that only particles with the spin $f$ and not also those of smaller spins could be assigned to the tensor field, and that  if the field $A_{ik...l}(x)$ describes a particle with spin $f$ satisfying the massive ``Schr\"{o}dinger-Gordon'' wave equation \eqref{SGeq}, then the number of linearly independent plane waves is $2f+1$, which differ by the orientation of the spin. Shortly after, in \cite{Fierz:1939ix} Fierz and Pauli applied the general formalism described in  \cite{Fierz:1939zz} to the particular case of spin $f=2$, simply taken as an example. Hence, they looked for the theory of a symmetric tensor field $A_{ij}(x)$ satisfying the massive wave equation 
\begin{equation}
\partial^2 A_{ij} = \kappa^2 A_{ij}\ ,
\label{SGeqspin2}\end{equation}
together with the secondary conditions
\begin{eqnarray}
A_{ii} &=& 0 \label{tracespin2}\\
\partial_i A_{ij} &=& 0\ .\label{transvspin2}
\end{eqnarray}
{\it Imposing the condition of tracelessness \eqref{tracespin2} by hand}, they 
derived the above equations \eqref{SGeqspin2} and \eqref{transvspin2} from the EOM of the following Lagrangian
\begin{equation}\label{FP lagrangian}
{\cal L}_{orig} = \kappa^2 A_{ij}A^{ij} + \partial_l A_{ij} \partial^l A^{ij} + a_1 \partial_i A^{ik} \partial_j {A^{j}}_{k} +  a_2 \kappa^2 C^2 + a_3 \partial_i C \partial^i C + \partial_i A^{ij} \partial_j C\; ,
\end{equation}
where an additional scalar field $C(x)$ is introduced, and $a_i,\  i=1,2,3$ are dimensionless constant parameters. They showed that the desired results \eqref{SGeqspin2} and \eqref{transvspin2} are obtained if
\begin{align}
a_1 = - 2 \; ; \\[8pt]
2a_3 = a_2 = - \frac{3}{4}\;.
\end{align}
In \eqref{FP lagrangian} the introduction of the scalar field $C(x)$ is an artifice which enables one to derive the transversality condition \eqref{transvspin2}. In a more modern language, we would call this scalar field a Nakanishi-Lautrup Lagrange multiplier \cite{Lautrup:1967zz,Nakanishi:1966zz} introduced to implement the gauge condition \eqref{transvspin2}. It is therefore interesting (and surprising !) to remark that, in the original article \cite{Fierz:1939ix}, Fierz and Pauli wrote an action for the graviton which included both a mass term and a gauge fixing term, implemented by means of the Lagrange multiplier $C(x)$.
It appears that the problem of treating the mass term as a gauge fixing  was absent in the original formulation. Moreover, the massless sector of the lagrangian $L_{orig}$ \eqref{FP lagrangian} does not coincide with the LG action $S_{LG}$ \eqref{SLG}, since the trace $A_{ii}(x)$ (a.k.a. $h(x)$) is set to zero {\it a priori}. As a consequence, the only mass parameter appearing in \eqref{FP lagrangian} is the one coupled to $A_{ij}A^{ij}$ (a.k.a. $h_{\mu\nu}h^{\mu\nu}$), which therefore corresponds to $m_1^2$ in \eqref{Sm}. 
It seems that, despite the fact that the original FP approach to LMG was realized in term of a gauge fixed  action, later developments left the gauge fixing term behind, leaving room for the divergent massless limit and for the vDVZ discontinuity. \\

In this paper we adopt a conservative policy, recovering the original FP approach to LMG, with a few important differences.

\subsection{Adding masses to the gauge fixed action}

One of the reasons of the divergent massless limit of the FP theory is that the mass term $S_m$ \eqref{Sm} serves as gauge fixing as well, in the sense that its presence is necessary to define the propagator \eqref{FPprop}, as shown in Section 2.2. The standard way to proceed in gauge field theory, instead, is first to gauge fix the invariant action, in order to have a well defined partition function $Z[J]$. This latter generates all the correlation functions of the theory, starting from the 2-points green function, a.k.a. the propagator. Only after having obtained that, the fields of the theory can be given masses through various procedures. In LMG this is achieved by adding to the gauge fixed action the mass term $S_m$ \eqref{Sm} \cite{Blasi:2017pkk}. So, let us first proceed by gauge fixing the invariant action $S_{LG}[h]$ \eqref{SLG}. The gauge field is represented by a rank-2 symmetric tensor $h_{\mu\nu}(x)$. Hence, the most general covariant gauge fixing is

\begin{equation}\label{gaugecond}
\partial_\mu h^{\mu \nu} + \kappa \partial^{\nu} h = 0\ ,
\end{equation}
which, by means of the usual Faddeev-Popov ($\Phi\Pi$) exponentiation \cite{Faddeev:1967fc}, yields the gauge fixing action term
\begin{equation}\label{Sgf}
    S_{gf}[h;k,\kappa]= -\frac{1}{2k} \int \mathrm{d^4 x} \ \Big[  \partial_\mu h^{\mu \nu} + \kappa \partial^{\nu} h  \Big]^2 \; .
\end{equation}
Notice that the gauge fixing term \eqref{Sgf} depends on two gauge parameters $k$ and $\kappa$, which play different roles. In fact, $k$ determines how the gauge fixing condition \eqref{gaugecond} is enforced. It can be seen as a kind of $primary$ gauge fixing parameter, which corresponds to the standard gauge fixing parameter of Yang-Mills theory: $k=0$ corresponds to the Landau gauge, for instance. On the other hand, the parameter $\kappa$ fine-tunes the class of gauge fixing identified by $k$. It plays a $secondary$ role. As an example, the harmonic, or Lorenz, gauge is obtained with the choice $\kappa=-1/2$. Hence, it makes sense to talk about harmonic-Landau gauge, for instance, meaning by that the choice $k=0$ and $\kappa=-1/2$. Once the action $S_{LG}[h]$ \eqref{SLG} has been gauge fixed by the gauge fixing term \eqref{Sgf}, we can add the mass term \eqref{Sm}, so that our starting point for a theory of LMG  is given by the action 

\begin{equation}\label{SMG}
S_{LMG} = S_{LG} + S_{gf} + S_{m} \; .
\end{equation}
The action \eqref{SMG} is the starting, rather than the arrival, point, because the road ahead of us is still long. We have indeed to face the problems which affect the FP theory: in particular the massless limit and the absence of the vDVZ discontinuity. 
But, first, we have to deal with the main feature of LMG: the five DOF which must characterize a spin-2 massive particle. Our comparison is the FP theory, which reaches this goal with one mass parameter only, because of the FP tuning \eqref{FPT}. The action \eqref{SMG}, instead, depends on two masses, for now.
This will be done in the next Section.

\section{Degrees of freedom}

A realistic theory of MG needs five propagating massive DOF. The easiest way to see this is to notice that the massive graviton is a spin $S=2$ particle, which displays $2S+1$ independent components. Hence, given that the graviton is described by a symmetric rank-2 tensor $h_{\mu\nu}(x)$, only five out of its ten components correspond to physical DOF. Therefore, a necessary condition for a gauge theory of a massive rank-2 symmetric tensor to be promoted to a theory of LMG is to recover the five linear equations represented by the constraints of tracelessness \eqref{tracespin2} (or \eqref{C2}) and of transversality \eqref{transvspin2} (or \eqref{C1}), in order to lower the number of independent components of $h_{\mu\nu}(x)$ from ten to five.
The realization of this necessary condition, in the framework of a well defined gauge field theory of a symmetric rank-2 tensor, is the main aim of this paper. 
To reach our goal, we shall now consider the EOM of the action \eqref{SMG} and we shall manage to restrict the mass parameters $m^2_1$ and $m^2_2$ to the cases in which we can find enough constraints to ensure the propagation of five massive DOF. Moreover, 
in order to deal with a physically consistent theory, we must also require that the propagating DOF do not depend on the gauge parameters $k$ and $\kappa$, allowing instead, of course, a  dependence on the mass parameters $m^2_1$ and $m^2_2$.\\
The action $S_{LMG}$ \eqref{SMG} in momentum space reads: 
\begin{equation} \label{quadr_action_mass}
    S_{LMG}= \int \mathrm{d^4 p} \ \tilde{h}_{\mu \nu} \; \Omega^{\mu \nu, \alpha \beta} \; \tilde{h}_{\alpha \beta}\; ,
\end{equation}
where the kinetic operator $\Omega$ is
\begin{multline}\label{kinetic_tensor}
   \Omega_{\mu \nu,\alpha \beta} = \frac{1}{2} \left[ m_2^2 - \left( 1 + \frac{\kappa^2}{k} \right)p^2 \right]\eta_{\mu \nu}\eta_{\alpha \beta} + \frac{1}{2}\left(1-\frac{\kappa}{k} \right)\left( \eta_{\mu \nu} e_{\alpha \beta} + \eta_{\alpha \beta} e_{\mu \nu}  \right)p^2 + \\[10pt]
    \frac{1}{2}(p^2 + m_1^2)\mathcal{I}_{\mu \nu,\alpha \beta} - \frac{1}{4}\left( 1 + \frac{1}{2k} \right)  \left( e_{\mu \alpha} \eta_{\nu \beta} + e_{\nu \alpha} \eta_{\mu \beta} +e_{\mu \beta} \eta_{\nu \alpha} +e_{\nu \beta} \eta_{\mu \alpha}\right) p^2\; ,
\end{multline}
$\mathcal{I}$ is the rank-4 tensor identity
 \begin{equation}
            \mathcal{I}_{\mu \nu, \rho \sigma} = \frac{1}{2} (\eta_{\mu \rho} \eta_{\nu \sigma} + \eta_{\mu \sigma} \eta_{\nu \rho}) 
\label{identity} \end{equation}

and $e_{\mu\nu}$ is the transverse projector
\begin{equation}
e_{\mu\nu}=\frac{p_\mu p_\nu}{p^2}\ .
\label{transvproj}\end{equation}

From the action $S_{LMG}$ \eqref{quadr_action_mass} we get the momentum space EOM

\begin{align}
      \frac{\delta S}{\delta \tilde{h}_{\mu \nu}} = \;& -\left( 1 +\frac{\kappa^2}{k}\right)\eta^{\mu \nu} p^2 \tilde{h} \;+\; \left( 1 -\frac{\kappa}{k}\right)p^\mu p^\nu \tilde{h} \;+\; \left( 1 -\frac{\kappa}{k}\right) \eta^{\mu \nu} p^\alpha p^\beta \tilde{h}_{\alpha \beta} \; \nonumber \\[10pt]
       & + p^2 \tilde{h}^{\mu \nu} -\left( 1 +\frac{1}{2k} \right)\big(p^{\mu} p^\alpha {\tilde{h}^\nu}_\alpha + p^{\nu} p^\alpha {\tilde{h}^\mu}_\alpha\big) \;+\; m_1^2 \, \tilde{h}^{\mu \nu} + m_2^2 \, \eta^{\mu \nu}\tilde{h}=0 \; . \label{quadr_eom}
\end{align}
In order to study the propagating DOF, we saturate the EOM \eqref{quadr_eom} with  $\eta_{\mu \nu}$ and $e_{\mu \nu}$ \eqref{transvproj}, to get

\begin{align}
    \boldsymbol{\eta_{\mu \nu}}:& \ \left[ (m_1^2 + 4 m_2^2) - \left( 2 + \frac{\kappa}{k} (1+ 4 \kappa) \right) p^2 \right] \: \tilde{h} \; + \; \left( 2 - \frac{1}{k}(1+4\kappa)\right)\:p^2 \: e_{\mu \nu} \tilde{h}^{\mu \nu} = 0 \label{DOF _eq_1_}\\[15pt]
    \boldsymbol{e_{\mu \nu}}:& \ \left[  m_2^2 -  \frac{\kappa}{k} (1+  \kappa) p^2 \right] \: \tilde{h} \; + \; \left[m_1^2 - \frac{1}{k}(1+\kappa)p^2 \right]\: e_{\mu \nu}\tilde{h}^{\mu \nu} = 0  \; . \label{DOF _eq_2_}
\end{align}

From these two equations we deduce that, if $m^2_1 \neq 0$ (we will consider the case $m^2_1=0$ later), the only solution is 
\begin{empheq}[left=\empheqlbrace]{align}
    \tilde{h} &= 0  \label{h_0_quadr}\\[10pt]
   e^{\mu \nu} \tilde{h}_{\mu \nu}&=  0  \label{eh}\; ,
\end{empheq}
since \eqref{DOF _eq_1_} and \eqref{DOF _eq_2_} form a homogeneous system of two linear equations which has a non trivial solution only if the determinant of the coefficients matrix vanishes. For $m^2_1 \neq	0$ this determinant cannot vanish, independently of the choice of $\kappa$, $k$ and $m_2^2$.
Substituting \eqref{h_0_quadr} and \eqref{eh}  into the EOM \eqref{quadr_eom} we get

\begin{equation}\label{quadr_almost_prop_m1}
    (p^2 + m_1^2) \tilde{h}_{\mu \nu} - \left( 1 + \frac{1}{2k} \right) \Big( p_{\mu} p^\alpha \tilde{h}_{\nu \alpha} + p_{\nu} p^\alpha \tilde{h}_{\mu \alpha} \Big) = 0 \; ,
\end{equation}

which, saturated with $p^\nu$ and using again \eqref{eh}, yields

\begin{equation}\label{ph}
    \left( m_1^2 - \frac{1}{2k}p^2 \right) p^\nu \tilde{h}_{\mu \nu}=0 \; .
\end{equation}

The above equation is satisfied if
\begin{equation}
p^2=2km_1^2
\label{2km1}\end{equation}
or
\begin{equation}
    p^\nu \tilde{h}_{\mu \nu}=0 \; ,
\label{phtilde=0}\end{equation}
but the requirement that the physical masses should not depend on the gauge parameters, implies that the only allowable solution of \eqref{ph} is \eqref{phtilde=0}.
This, together with \eqref{h_0_quadr}, gives the five constraints
\begin{empheq}[left=\empheqlbrace]{align}
    \tilde{h} &= 0  \label{quadr_constraint_1}\\[10pt]
   p^\mu \tilde{h}_{\mu \nu}&=  0 \; , \label{quadr_constraint_2}
\end{empheq}

which ensure the propagation of five DOF. Conditions \eqref{quadr_constraint_1} and \eqref{quadr_constraint_2} inserted into the EOM \eqref{quadr_eom} give the massive propagation of $h_{\mu \nu}$
\begin{equation}\label{quadr_prop_m1}
    (p^2 + m_1^2) \tilde{h}_{\mu \nu}=0 \; ,
\end{equation}

which, interestingly, does not depend on the mass parameter $m_2$. 
If, on the other hand, $m^2_1=0$, the two equations \eqref{DOF _eq_1_} and \eqref{DOF _eq_2_} have non trivial solutions only if 
\begin{equation}
    \kappa=\,-1 \ \ ;\ \ 
    k=\, -\frac{3}{2} \; ,
\end{equation}
which, plugged back into \eqref{DOF _eq_1_} and \eqref{DOF _eq_2_}  with $m^2_1=0$, yield
\begin{equation}
    h=0\; .
\end{equation}

Substituting $m^2_1=0$ and $h=0$ in the EOM \eqref{quadr_eom} we notice that all the dependence on the mass parameters vanishes. 
The resulting EOM would therefore describe the propagation of a massless field or, alternatively, the propagation of a field with a mass dependent on the gauge parameters $k$ and $\kappa$. Both cases do not represent acceptable descriptions of a LMG theory, therefore we conclude that it must be
\begin{equation}
m_1^2\neq 0\ .
\label{m1nonzero}\end{equation}
To summarize, we showed that the gauge fixed action $S_{LMG}$ \eqref{SMG}, with the mass term $S_{m}$ \eqref{Sm}, displays five DOF, provided that $m^2_1\neq0$, for any value of $m_2^2$ and for arbitrary gauge parameters $k$ and $\kappa$.

\section{Propagators}

The momentum space propagator $G_{\mu \nu, \alpha \beta} (p)$
is defined by the condition

\begin{equation}\label{prop_def}
    {\Omega_{\mu \nu}}^{\alpha \beta}G_{\alpha \beta, \rho \sigma} = \mathcal{I}_{\mu \nu, \rho \sigma} \; ,
\end{equation}

where $\Omega_{\mu \nu,\alpha \beta}(p)$ is the kinetic operator introduced in \eqref{kinetic_tensor} and $ \mathcal{I}_{\mu \nu, \rho \sigma}$ is the rank-4 tensor identity \eqref{identity}.
In order to find $G_{\mu \nu, \alpha \beta}(p)$ it is convenient to introduce the rank-2 projectors $e_{\mu \nu}$ \eqref{transvproj} and

\begin{equation}\label{dmunu}
d_{\mu \nu} \equiv \eta_{\mu \nu} -  e_{\mu \nu }
\end{equation}

which are idempotent and orthogonal

\begin{equation}
    e_{\mu \lambda} {e^\lambda}_\mu = e_{\mu \nu}, \ \mathrm{}\   d_{\mu \lambda} {d^\lambda}_\nu = d_{\mu \nu}, \ \mathrm{}\     e_{\mu \lambda} {d ^\lambda}_\nu  =0 \: .  
\label{defed}\end{equation}

With these rank-2 projectors we can construct a basis formed by five rank-4 tensors which we collectively denote
\begin{equation}
X_{\mu \nu, \alpha \beta} \equiv (A,B,C,D,E)_{\mu\nu,\alpha\beta}
\label{defX}\end{equation}
 with the symmetry properties 

\begin{equation}
X_{\mu \nu, \alpha \beta}  = X_{\nu \mu, \alpha \beta}     = X_{\mu \nu, \beta \alpha }    = X_{\alpha \beta, \mu \nu}      \; .
\end{equation}

In terms of the projectors $e_{\mu\nu}$ and $d_{\mu\nu}$ the operators $X_{\mu \nu, \alpha \beta}$ read
\begin{align}
    A_{\mu \nu, \alpha \beta} &= \frac{d_{\mu \nu} d_{\alpha \beta}}{3} \label{A}\\[10pt]
 B_{\mu \nu, \alpha \beta} &= e_{\mu \nu} e_{\alpha \beta} \label{B}\\[10pt]
  C_{\mu \nu, \alpha \beta} &= \frac{1}{2} \left(  d_{\mu \alpha} d_{\nu \beta} + d_{\mu \beta} d_{\nu \alpha} - \frac{2}{3} d_{\mu \nu} d_{\alpha \beta}  \right) \label{C}\\[10pt]
  D_{\mu \nu, \alpha \beta} &=  \frac{1}{2} \left(  d_{\mu \alpha} e_{\nu \beta} + d_{\mu \beta} e_{\nu \alpha} + e_{\mu \alpha} d_{\nu \beta} + e_{\mu \beta} d_{\nu \alpha}  \right)\label{D}\\[10pt]
  E_{\mu \nu, \alpha \beta} &= \frac{\eta_{\mu \nu} \eta_{\alpha \beta}}{4} \label{E} \; ,
\end{align}

and have the following properties:

\begin{itemize}
    \item 
    decomposition of the identity $\mathcal{I}_{\mu \nu, \alpha\beta}$ \eqref{identity}~:
    \begin{equation}
    A_{\mu \nu , \alpha \beta} + B_{\mu \nu , \alpha \beta} + C_{\mu \nu , \alpha \beta} + D_{\mu \nu , \alpha \beta} =  \mathcal{I}_{\mu \nu , \alpha \beta}\ ;
    \label{idempotency}\end{equation}
    \item idempotency~:
    \begin{equation}
    X_{\mu\nu}^{\ \ \rho\sigma}X_{\rho\sigma,\alpha\beta}=X_{\mu\nu,\alpha\beta}\ ;
    \label{idempotency}\end{equation}
    \item orthogonality of $A$, $B$, $C$ and $D$~:
    \begin{equation}
    X_{\mu \nu , \alpha \beta} {{X^\prime}^{\alpha \beta}}_{\rho \sigma} = 0\ \ \mbox{if}\
    (X,X^\prime)\neq E\ \mbox{and}\ X\neq X^\prime\ ;
    \label{orthogonality}\end{equation}
    \item contractions with $E$~:
    \begin{align} 
    A_{\mu \nu , \alpha \beta} {E^{\alpha \beta}}_{\rho \sigma} 
    &= \frac{d^{\mu \nu} \eta_{\rho \sigma}}{4}\label{AE}\\[10pt] 
    B_{\mu \nu , \alpha \beta} {E^{\alpha \beta}}_{\rho \sigma} 
    &= \frac{e^{\mu \nu} \eta_{\rho \sigma}}{4}\label{BE}\\[10pt]
    C_{\mu \nu , \alpha \beta} {E^{\alpha \beta}}_{\rho \sigma}
    &=D_{\mu \nu , \alpha \beta} {E^{\alpha \beta}}_{\rho \sigma}=0\label{CE-DE}\ .    \end{align}
    
\end{itemize}

The kinetic operator $\Omega_{\mu \nu , \alpha \beta}(p)$ \eqref{kinetic_tensor} can be written in terms of the rank-4 projectors \eqref{A}-\eqref{E}~:

\begin{equation}
    \Omega_{\mu \nu, \alpha \beta} = t A_{\mu \nu, \alpha \beta} + u B_{\mu \nu, \alpha \beta} + v C_{\mu \nu, \alpha \beta} + z D_{\mu \nu, \alpha \beta} + w E_{\mu \nu, \alpha \beta} \; ,
\end{equation}

where, after a lenghty but straightforward calculation, the coefficients are given by

{\small
\begin{align}
    t&= \left( \frac{3\kappa}{2k} -1\right) p^2 + \frac{1}{2}m_1^2 \\[5pt]
    u&= -\frac{1}{4k}\left( 2\kappa + 1\right) p^2 + \frac{1}{2}m_1^2\\[5pt]
    v&= \frac{1}{2}(p^2 + m_1^2)\\[5pt]
    z&= -\frac{1}{4k} p^2 + \frac{1}{2}m_1^2\\[5pt]
    w&=- \frac{2\kappa}{k}(1+\kappa)p^2 + 2 m_2^2 \; .
\end{align}
}
Similarly, we can expand the propagator $G_{\mu \nu , \alpha \beta}(p)$~:
 
\begin{equation}\label{QUADR_propagator}
     G_{\mu \nu, \alpha \beta} =\text{ }\hat{t} A_{\mu \nu, \alpha \beta} + \hat{u} B_{\mu \nu, \alpha \beta} + \hat{v} C_{\mu \nu, \alpha \beta} + \hat{z} D_{\mu \nu, \alpha \beta} + \hat{w} E_{\mu \nu, \alpha \beta} \; ,
\end{equation}
where $\hat{t}$, $\hat{u}$, $\hat{v}$, $\hat{z}$ and $\hat{w}$ are functions of the momentum $p$ and depend on the gauge parameters $k$ and $\kappa$ appearing in $S_{gf}$ \eqref{Sgf} and on the masses $m^2_1$ and $m^2_2$ of the mass term $S_m$ \eqref{Sm}. 
Solving the defining equation \eqref{prop_def}, we find~:

\begin{align}
    \hat{t}=&\; \frac{2(1+\kappa)(1+ 4 \kappa) p^2 - 2k(m_1^2 + 4m_2^2)}{\text{DN}(m_1,m_2,k,\kappa,p^2)} \label{quadr_coeff_1}\\[10pt]
    \hat{u}=&\; \frac{2\big[\kappa (1+4\kappa) +2k\big]p^2 - 2k(m_1^2 + 4m_2^2)}{\text{DN}(m_1,m_2,k,\kappa,p^2)}\\[10pt]
    \hat{v}=&\; \frac{2}{p^2 + m_1^2} \label{quadr_coeff_3}\\[10pt]
    \hat{z}=&\; \frac{-4k}{p^2-2km_1^2}\\[10pt]
    \hat{w}=&\; \frac{8km_2^2 - 8\kappa(1+\kappa) p^2}{\text{DN}(m_1,m_2,k,\kappa,p^2)}   \; , \label{quadr_coeff_5}
\end{align}
\\[10pt]
\normalcolor
where we used the shorthand notation for the denominator
\begin{align}
    \text{DN}&(m_1,m_2,k,\kappa,p^2) \equiv \nonumber \\[10pt]
    &-2(1+\kappa)^2 p^4 + \Big[( 1+2\kappa + 4\kappa^2 + 2k)m_1^2 + (3+2k)m_2^2\Big] p^2 -4 k m_1^2 m_2^2 - k m_1^4\; .  \label{denominator}
\end{align} 
The propagator \eqref{QUADR_propagator} displays poles that depend on the gauge parameters $k$ and $\kappa$ and might even be tachyonic. This does not come as a surprise, given that the theory describes the dynamics of a rank-2 symmetric tensor field, and the pole structure of its propagator is  more complicated than the usual scalar, spinor or vector cases. Neither it should be seen as a problem, since in the previous Section we proved that only five of the ten components of $h_{\mu \nu}$ represent independent DOF, which satisfy the massive wave equation of the Klein-Gordon type \eqref{quadr_prop_m1} which depends on the mass $m_1^2$ only, in agreement with the FP theory. Hence, looking at the whole propagator is neither helpful nor correct in order to identify the physical pole in this case, since we may allow for non physical poles located in non physical sectors of the propagator.
Rather, what should be done in order to select the physical pole, is to look to the observables related to the propagator \eqref{QUADR_propagator}. An important 
example is the one already considered in the analysis of the vDVZ discontinuity: the scattering amplitude of light and a massive object, mediated by the gravitational interaction, which is responsible for the observed time delay in gravitational lensing. Formally, this observable can be traced back to the more general interaction amplitude of two conserved energy-momentum tensors $T^{(1)}_{\mu \nu}$ and $T^{(2)}_{\mu \nu}$, of which the one corresponding to light (which in the following we choose to be  $T^{(2)}_{\mu \nu}$) is traceless~: 
\begin{equation}\label{scattering amplitude}
    \tilde T^{(1)}_{\mu \nu} G^{\mu \nu, \alpha \beta} \tilde T^{(2)}_{\alpha \beta}\; .
\end{equation}  
Substituting the propagator \eqref{QUADR_propagator} into \eqref{scattering amplitude}, we get
\begin{equation}\label{scattering amplitude_2}
 \tilde T^{(1)}_{\mu \nu} \Big(  \hat{t} A^{\mu \nu, \alpha \beta} + \hat{u} B^{\mu \nu, \alpha \beta} + \hat{v} C^{\mu \nu, \alpha \beta} + \hat{z} D^{\mu \nu, \alpha \beta} + \hat{w} E^{\mu \nu, \alpha \beta} \Big) \tilde T^{(2)}_{\alpha \beta}\ .
\end{equation}

The above expression contains contractions of the tensor projectors \eqref{A}-\eqref{E} with $\tilde T^{(1)}_{\mu \nu}$ and $\tilde T^{(2)}_{\mu \nu}$. These are greatly simplified thanks to the fact that the energy-momentum tensors are conserved: $p^\nu \tilde{T}_{\mu \nu}^{(1)} =p^\nu \tilde{T}_{\mu \nu}^{(2)}= 0$. In particular: 

\begin{align}
& \tilde T^{(1)}_{\mu \nu}  A^{\mu \nu, \alpha \beta}\tilde T^{(2)}_{\alpha \beta} = \frac{1}{3}  \tilde T^{(1)}_{\mu \nu}  \eta^{\mu \nu} \eta^{\alpha \beta}\tilde T^{(2)}_{\alpha \beta} \label{simp1}\\[8pt]
& \tilde T^{(1)}_{\mu \nu}  B^{\mu \nu, \alpha \beta}\tilde T^{(2)}_{\alpha \beta} = 0\\[8pt]
& \tilde T^{(1)}_{\mu \nu}  C^{\mu \nu, \alpha \beta}\tilde T^{(2)}_{\alpha \beta} = \frac{1}{2}  \tilde T^{(1)}_{\mu \nu} ( \eta^{\mu \alpha} \eta^{\nu \beta} + \eta^{\mu \beta} \eta^{\nu \alpha} - \frac{2}{3}\eta^{\mu \nu} \eta^{\alpha \beta} )\tilde T^{(2)}_{\alpha \beta}\\[8pt]
& \tilde T^{(1)}_{\mu \nu}  D^{\mu \nu, \alpha \beta}\tilde T^{(2)}_{\alpha \beta} = 0\\[8pt]
& \tilde T^{(1)}_{\mu \nu}  E^{\mu \nu, \alpha \beta}\tilde T^{(2)}_{\alpha \beta} = \frac{1}{4}  \tilde T^{(1)}_{\mu \nu}  \eta^{\mu \nu} \eta^{\alpha \beta}\tilde T^{(2)}_{\alpha \beta} \label{simp5}\ .
\end{align}

Eqs. \eqref{simp1}-\eqref{simp5} can be further simplified by using the fact that $\tilde{T}_{\mu \nu}^{(2)}$ is traceless ($\eta^{\mu\nu} \tilde{T}_{\mu \nu}^{(2)}= 0$), which means that the only non-vanishing contraction left is 

\begin{equation}
\tilde T^{(1)}_{\mu \nu}  C^{\mu \nu, \alpha \beta}\tilde T^{(2)}_{\alpha \beta} = \frac{1}{2}  \tilde T^{(1)}_{\mu \nu} ( \eta^{\mu \alpha} \eta^{\nu \beta} + \eta^{\mu \beta} \eta^{\nu \alpha} )\tilde T^{(2)}_{\alpha \beta}\ .
\end{equation}

Therefore the scattering amplitude \eqref{scattering amplitude} reduces to

\begin{equation}\label{final_amp_v}
   \tilde  T^{(1)}_{\mu \nu} \Big(  \hat{v}\; \mathcal{I}^{\mu \nu, \alpha \beta}  \Big) \tilde T^{(2)}_{\alpha \beta}\; ,
\end{equation} 
according to which only the pole contained in the coefficient $\hat{v}$ \eqref{quadr_coeff_3} plays a physical role. Reassuringly, that pole is 
\begin{equation}\label{real mass}
p^2 = - m_1^2\ ,
\end{equation}
which confirms that the theory describes a graviton with mass $m^2_1$, in agreement with the massive wave equation \eqref{quadr_prop_m1}, which was obtained by studying the EOM deriving from the action $S_{LMG}$ \eqref{SMG}, and also with the FP theory, \eqref{K-G}. Notice that the same argument may be repeated to show that also the scattering amplitude between two radiation-like objects, both described by conserved and traceless energy-momentum tensors, isolates \eqref{real mass} as the unique physical pole of the propagator \eqref{QUADR_propagator}.

\section{Absence of the vDVZ discontinuity}

In Section 2.2 we reviewed the vDVZ discontinuity, which concerns the mismatch between the gravitational interaction of two energy-momentum tensors in the pure massless theory described by the action $S_{LG}$ \eqref{SLG}  and in the massless limit of FP theory $S_{FP}$ \eqref{SFP}. In \cite{Gambuti:2020onb} it has been shown that in a particular gauge (namely the $k=-\frac{1}{2}$, $\kappa=-\frac{1}{2}$ harmonic one) the theory described by the action $S_{LMG}$ \eqref{SMG} is free of the vDVZ discontinuity. In this Section we generalize the result of \cite{Gambuti:2020onb} to all possible gauge choices, $i.e.$ for every value of the gauge parameters $k$ and $\kappa$. 
With the propagator $G_{\mu\nu,\alpha\beta}$ \eqref{QUADR_propagator} derived from the gauge fixed action $S_{LMG}$ \eqref{SMG}, hence for generic $k$ and $\kappa$ gauge parameters,   we can compute the interaction amplitudes between two non-relativistic conserved energy-momentum tensors ${T_{\mu \nu}^{(1)}}$ and ${T_{\mu \nu}^{(2)}}$ and compare them to the corresponding LG amplitudes,  \eqref{INT1} and \eqref{INT2}. As explained in Section 2.2, the interaction of the graviton field with an external energy-momentum tensor is obtained by means of the interaction term $S_{int}$ \eqref{int term}. To distinguish the gauge fixed theory $S_{LMG}$ \eqref{SMG} from the LG and FP theories, we denote its coupling constant with $\lambda_{LMG}$. 
Following the steps described in \cite{Rubakov:2008nh}, if neither ${T_{\mu \nu}^{(1)}}$ nor ${T_{\mu \nu}^{(2)}}$ are traceless, which corresponds to the scattering between two massive bodies, the resulting amplitude in the massless limit ($m^2_1, m^2_2) \rightarrow 0$ of the theory is

\begin{equation}\label{INT1_GF}
\lambda^2_{LMG}\: \tilde{T}^{(1)}_{\mu \nu} G^{\mu \nu, \alpha \beta} \tilde{T}^{(2)}_{\alpha \beta} = \lambda^2_{LMG}\: \tilde{T}^{(1)}_{00} \tilde{T}^{(2)}_{00} \frac{1}{p^2}\; ,
\end{equation}

whereas, if ${T_{\mu \nu}^{(2)}}$ is traceless, corresponding to the scattering between light and a massive body, like in the gravitational lensing, we have

\begin{equation}\label{INT2_GF}
\lambda^2_{LMG}\: \tilde{T}_{\mu \nu}^{(1)} G^{\mu \nu, \alpha \beta} \tilde{T}^{(2)}_{\alpha \beta} = \lambda^2_{LMG}\: \tilde{T}^{(1)}_{00} \tilde{T}^{(2)}_{00} \frac{2}{p^2}\; .
\end{equation}

It is straightforward to see that if we set 

\begin{equation}
\lambda_{LMG} = \lambda_{LG}
\end{equation}

the LG amplitudes \eqref{INT1} and \eqref{INT2} exactly match \eqref{INT1_GF} and \eqref{INT2_GF} respectively, which proves the absence of the vDVZ discontinuity {\it in a gauge independent way}.
We stress that, although the massive propagator $G_{\mu \nu, \alpha \beta}$ \eqref{QUADR_propagator} depends on the gauge parameters $k$ and $\kappa$ through the coefficients \eqref{quadr_coeff_1}-\eqref{quadr_coeff_5}, the massless limit of the interaction amplitudes \eqref{INT1_GF} and \eqref{INT2_GF} is gauge independent and coincides with the LG prediction. This is related to the fact that the massless limit of the action $S_{LMG}$ \eqref{SMG} is the LG gauge fixed action 

\begin{equation}
S_{LG} + S_{gf}\ ,
\end{equation}
where $S_{LG}$ and $S_{gf}$ are given by \eqref{SLG} and \eqref{Sgf}, respectively.
Hence, in the massless limit, the propagators of the two theories coincide. This is the basic reason of the absence of the vDVZ discontinuity. 
In other words, the vDVZ discontinuity is a direct consequence of the structure of the FP action, where the mass term $S_m$ \eqref{Sm} is added directly to the invariant action $S_{LG}$ \eqref{SLG}, therefore acting effectively as a gauge fixing term. Without the mass term, the invariant action does not have a propagator. Therefore, the mass parameters, which also play the role of gauge fixing parameters, cannot be physical. In the massless limit, the FP theory is not dynamical ($i.e.$ does not have a propagator), and therefore it is not surprising that a discontinuity arises.}
Throughout this paper, we repeatedly claimed that 
the vDVZ discontinuity may be related to the
nature of the FP mass term as a gauge fixing term.  The massive gravity propagator in \eqref{FPprop} is
not well defined in the zero mass limit due to the lack of gauge fixing
in the massless theory. However, the vDZV discontinuity appears in the
analysis of the propagator contracted with energy-momentum tensors like
in \eqref{scattering amplitude}. This quantity is gauge invariant in linearized GR
(for conserved sources). Hence one might expect that the ill defined terms in
the zero mass limit of massive FP propagator drop out of equations like
\eqref{scattering amplitude} which contribute to the vDVZ discontinuity (this quantity indeed
remains finite in the zero mass limit). 
The apparent contradiction between this observation and the explanation
provided in this paper is explained by observing that when we say that the mass term $S_{m}$ \eqref{Sm} at the FP point \eqref{FPT} has the role of a gauge fixing in the FP theory,  we are only referring to the fact that it enables us to invert the quadratic part of the action.  In fact,  not all quadratic terms which make it possible to find a propagator are also legitimate gauge fixing terms.  The gauge fixing procedure is much more than this. Indeed, it consists in limiting the functional integral in order to choose one representative  for each gauge orbit (modulo Gribov copies). This is realized through the highly non trivial $\Phi\Pi$ trick.  In the case of an abelian theory such as linearized gravity,  where the ghost sector decouples,  the $\Phi\Pi$ gauge fixing is realized by adding to the action the square of a functional $F^{\mu}[h_{\alpha \beta}]$,  where $F$ must carry the same number of Lorentz indices as the gauge parameter of the theory,  defined in our case in \eqref{gaugesym}.  Since the mass term cannot be written as the square of any such functional,  it follows that $S_{m}$ is not a true gauge fixing term and therefore no contradiction arises when comparing the linearized GR and FP propagators.  

\section{Summary of results and Discussion} 

In this paper we presented a gauge field theory of a massive symmetric tensor field describing a massive spin-2 particle. We summarize here the main features which allow to interpret this theory as an alternative to the FP theory of LMG.
\begin{enumerate}
\item \textbf{finite massless limit}\\
	The theory described by the action $S_{LMG}$ \eqref{SMG} yields the propagator 		$G_{\mu\nu,\alpha\beta}$ \eqref{QUADR_propagator} which is regular in the limits ($m_1^2,m_2^2)\rightarrow0$\ ;
\item \textbf{five massive DOF}\\
	In Section 3 we showed that only five of the ten components of the massive symmetric 	tensor $h_{\mu\nu}(x)$ actually represent DOF, as required for a massive graviton\ .
	The remarkable fact is that the mass associated with the propagation of the five DOF is 	$m^2_1$, while the value of $m^2_2$ is irrelevant, under this respect. Although obtained in a quite different way, this result  is in 		agreement with the FP theory, which is characterized by one mass parameter only\ ;
 \item \textbf{gauge independence}\\
	In gauge field theory, any claim concerning physical properties should not depend on a 	particular gauge choice. In our case, this requirement is met in the determination of the 	physical DOF of the theory: none of the steps leading to the five constraints represented  by \eqref{quadr_constraint_1} and 	\eqref{quadr_constraint_2} rely on particular values of the gauge parameters $k$ and $\kappa$\ ;
\item \textbf{continuity with LG}\\
	Any candidate theory for LMG must display a finite massless limit, which implies a non 	divergent propagator, and, in this limit, it should also provide physical predictions in 		agreement with GR. Physically, LMG should represent only a small correction of GR, 		therefore the mass term \eqref{Sm} should be seen as a perturbation. 
	In particular, the effects of a mass of the graviton should become relevant only at very 	large distances, while at smaller scales GR predictions must be restored. As discussed in Section 2.2, a serious flaw of the FP theory of LMG is the vDVZ discontinuity \cite{vanDam:1970vg,Zakharov:1970cc}. This well known issue is fixed by the Vainshtein 	mechanism \cite{Hinterbichler:2011tt,Vainshtein:1972sx,deRham:2014zqa}, which recovers the non linearities of GR in order to shield the effect of the extra scalar DOF introduced by the St\"{u}ckelberg 	formalism \cite{Stueckelberg:1900zz} in the massless limit of the FP theory. Our gauge fixed theory of LMG restores the continuity with	GR in a more natural way, without the introduction of additional fields, cutoffs or non linearities of the theory. In \cite{Gambuti:2020onb} it has been shown that in the particular 
$(k=\kappa = - \frac{1}{2})$ harmonic gauge, the theory described by the action $S_{LMG}$ \eqref{SMG} does not display the vDVZ discontinuity. 	
In Section 5 this result has been generalized to any gauge choice.
The fact that the 	gauge fixed massive theory described by the action $S_{LMG}$ \eqref{SMG} is not affected by the vDVZ discontinuity with GR encourages to believe that 	the way of approaching LMG presented in this paper is correct. Moreover, although the absence of the vDVZ discontinuity 	is a specific test and it does not prove in general that every measurable quantity is 		continuous with GR, the way the theory itself was constructed might imply this result. In 	fact, the evaluation of any 	observable  in quantum field theory needs a gauge fixing in order to define a 	propagator, which is 	guaranteed by the $\Phi \Pi$ procedure, adopted in this paper.
	The key feature of our approach is that the massive action $S_{LMG}$ \eqref{SMG} in the zero-mass 	limit becomes \textit{exactly} the $\Phi \Pi$ gauge fixed LG action, $i.e.$ the one we 		would have used to calculate the propagator and every other quantity in massless LG as 	well. Therefore, while it does not come as a surprise that the gravitational coupling 		between non relativistic matter and light turns out to be continuous with GR,  we may also infer that every other observable is, indeed, continuous with LG, in the massless 	limit. Moreover, the striking difference between the FP theory of LMG, where the vDVZ 	discontinuity is present, and the massive $\Phi \Pi$ gauge fixed theory described in this 	paper, where it is absent, gives evidence of the fact that, as anticipated, the FP theory is 	not a sub-case of our theory despite the fact that our approach contains the FP 		tuning \eqref{FPT}\ ;
\item \textbf{role of the mass parameters}\\
We stress as a remarkable fact that, although by means of a completely different approach, we recover here the main result of the FP theory, $i.e.$ that only the mass parameter $m^2_1$ associated   to $h^{\mu\nu}h_{\mu\nu}$ in the action $S_{m}$ \eqref{Sm} appears in the propagation of the physical DOF through the massive wave equation \eqref{quadr_prop_m1}, which is the starting request \eqref{SGeqspin2} of the original FP theory \cite{Fierz:1939ix}. Therefore,  the  mass of the graviton should be identified by $m^2_1$.

\end{enumerate} 

\section*{Acknowledgments}
We gratefully acknowledge Gianluca Gemme for his phenomenological hints and enlightening observations.

\end{document}